\documentclass[10pt,journal]{IEEEtran}

\usepackage[subtle, bibbreaks=normal, paragraphs=tight, floats=tight, mathspacing=tight, wordspacing=normal, tracking=normal]{savetrees}
\usepackage{amsmath}
\usepackage{amssymb}
\usepackage{bm}
\usepackage{color}
\usepackage{graphicx, floatrow}
\usepackage{empheq}
\usepackage[linesnumbered,ruled,lined]{algorithm2e}
\usepackage[outdir=./]{epstopdf}
\usepackage{caption}
\usepackage{lipsum}
\usepackage{cite}
\usepackage{multirow}
\usepackage{hyperref}
\hypersetup{hidelinks,
	colorlinks=true,
	allcolors=black,
	pdfstartview=Fit,
	breaklinks=true}
\usepackage{subcaption}
\usepackage{fancyhdr}
\usepackage{graphicx}
\SetKwInput{Input}{input}
\SetKwInput{Output}{output}

\usepackage{array}
\newcolumntype{L}[1]{>{\raggedright\let\newline\\\arraybackslash\hspace{0pt}}m{#1}}
\newcolumntype{C}[1]{>{\centering\let\newline\\\arraybackslash\hspace{0pt}}m{#1}}
\newcolumntype{R}[1]{>{\raggedleft\let\newline\\\arraybackslash\hspace{0pt}}m{#1}}
\newtheorem{remark}{Remark}
\IEEEoverridecommandlockouts
\usepackage[acronym]{glossaries-extra}
\setabbreviationstyle[acronym]{long-short}
\newacronym{1dsca}{1D-SCA}{One-Dimensional Search Successive Convex Approximation}
\newacronym{3gpp}{3GPP}{3rd Generation Partnership Project}
\newacronym{awgn}{AWGN}{Additive White Gaussian Noise}
\newacronym{bc}{BC}{Broadcast Channel}
\newacronym{bler}{BLER}{Block Error Rate}
\newacronym{bs}{BS}{Base Station}
\newacronym{cnoma}{C-NOMA}{cooperative NOMA}
\newacronym{csir}{CSIR}{Channel State Information at the Receiver}
\newacronym{csit}{CSIT}{Channel State Information at the Transmitter}
\newacronym{csi}{CSI}{Channel State Information}
\newacronym{df}{DF}{Decode-and-Forward}
\newacronym{dof}{DoF}{Degree-of-Freedom}
\newacronym{embb}{eMBB}{Enhanced Mobile Broadband}
\newacronym{fbl}{FBL}{Finite Blocklength}
\newacronym{hd}{HD}{Half-Duplex}
\newacronym{iid}{i.i.d}{independent and identically distributed}
\newacronym{leo}{LEO}{Low Earth Orbit}
\newacronym{mimo}{MIMO}{Multiple-Input Multiple-Output}
\newacronym{miso}{MISO}{Multiple-Input Single-Output}
\newacronym{mmf}{MMF}{Max-Min Fairness}
\newacronym{mmtc}{mMTC}{massive Machine Type Communications}
\newacronym{mrt}{MRT}{Maximum Ratio Transmission}
\newacronym{ndf}{NDF}{Non-regenerative Decode-and-Forward}
\newacronym{noma}{NOMA}{Non-Orthogonal Multiple Access}
\newacronym{nr}{NR}{New Radio}
\newacronym{oma}{OMA}{Orthogonal Multiple Access}
\newacronym{rsma}{RSMA}{Rate-Splitting Multiple Access}
\newacronym{sca}{SCA}{Successive Convex Approximation}
\newacronym{sc}{SC}{Superposition Coding}
\newacronym{sdma}{SDMA}{Space Division Multiple Access}
\newacronym{sic}{SIC}{Successive Interference Cancellation}
\newacronym{simo}{SIMO}{Single-Input Multiple-Output}
\newacronym{sinr}{SINR}{Signal to Interference plus Noise Ratio}
\newacronym{siso}{SISO}{Single-Input Single-Output}
\newacronym{soc}{SOC}{Second Order Cone}
\newacronym{svd}{SVD}{Singular Value Decomposition}
\newacronym{uav}{UAV}{Unmanned Aerial Vehicle}
\newacronym{urllc}{URLLC}{Ultra-Reliable and Low-Latency Communications}
\newacronym{v2x}{V2X}{Vehicle-to-Everything}
\newacronym{wmmse}{WMMSE}{Weighted Minimum Mean Square Error}

\begin{document}
\title{Max-Min Fairness of Rate-Splitting Multiple Access with Finite Blocklength Communications}
\makeatletter
\def\ps@IEEEtitlepagestyle{
    \def\@oddfoot{\mycopyrightnotice}
    \def\@evenfoot{}
}
\def\mycopyrightnotice{
    {\footnotesize
            \begin{minipage}{\textwidth}
                \centering
                Copyright~\copyright~2015 IEEE. Personal use of this material is permitted. However, permission to use this \\
                material for any other purposes must be obtained from the IEEE by sending a request to pubs-permissions@ieee.org.
            \end{minipage}
        }
}
\author{

\IEEEauthorblockN{Yunnuo~Xu, Yijie~Mao, \IEEEmembership{Member, IEEE}, Onur~Dizdar, \IEEEmembership{Member, IEEE} and~Bruno~Clerckx, \IEEEmembership{Fellow, IEEE}\vspace{-2.5em}}
\thanks{(\textit{Corresponding author: Yijie Mao.})
\par Y. Xu is with Imperial College London, London SW7 2AZ, UK (email: {yunnuo.xu19}@imperial.ac.uk).
\par Y. Mao is with the School of Information Science and Technology, ShanghaiTech University, Shanghai 201210, China (email: maoyj@shanghaitech.edu.cn).
\par Onur Dizdar was with Imperial college london. He is now with Viavi Solutions Inc (email: onur.dizdar@viavisolutions.com).
\par B. Clerckx is with the Department of Electrical and Electronic Engineering at Imperial College London, London SW7 2AZ, UK and with Silicon Austria Labs (SAL), Graz A-8010, Austria (email: b.clerckx@imperial.ac.uk).
}
}
\maketitle

\begin{abstract}
    Rate-Splitting Multiple Access (RSMA) has emerged as a flexible and powerful framework for wireless networks. In this paper, we investigate the user fairness of downlink multi-antenna RSMA in short-packet communications with/without cooperative (user-relaying) transmission. We design optimal time allocation and linear precoders that maximize the Max-Min Fairness (MMF) rate with Finite Blocklength (FBL) constraints. The relation between the MMF rate and blocklength, as well as the impact of cooperative transmission are investigated. Numerical results demonstrate that RSMA can achieve the same MMF rate as Non-Orthogonal Multiple Access (NOMA) and Space Division Multiple Access (SDMA) with smaller blocklengths (and therefore lower latency), especially in cooperative transmission deployment. Hence, we conclude that RSMA is a promising multiple access for guaranteeing user fairness in low-latency communications.
\end{abstract}
\vspace{-1mm}

\vspace{-4mm}
\section{Introduction}
\vspace{-1mm}
During the past few years, there has been a consensus that 5G will support three scenarios, \gls{embb}, \gls{mmtc} and \gls{urllc}. 
According to the \gls{3gpp}, a general \gls{urllc} reliability requirement is $99.999\%$ with latency being less than $1$ ms \cite{2022}. To support low-latency communications, short-packet with \gls{fbl} codes are adopted to reduce the transmission delay \cite{7529226}. 
\par In contrast to Shannon's capacity with infinite blocklength assumption, \gls{bler} cannot be neglected in the \gls{fbl} regime \cite{7529226}. Polyanskiy et al. provided information-theoretic limits on the achievable rate for given \gls{fbl} and \gls{bler} in \gls{awgn} fading channels \cite{5452208}. The authors in \cite{6034120} extended the achievable rate expression to \gls{siso} stationary fading channels with perfect \gls{csi}. The maximum achievable transmission rate was investigated over  
\gls{mimo} fading channels under both perfect and imperfect \gls{csi} settings \cite{6802432}.
\par It is known that \gls{noma} outperforms \gls{oma} by employing \gls{sc} at transmitter and \gls{sic} at receiver. The superiority of \gls{noma} over \gls{oma} was shown in terms of capacity region under infinite blocklength assumption \cite{1054727}, and \gls{mmf} throughput with \gls{fbl} constraints \cite{9482456}. With cooperative transmission, \gls{cnoma} can achieve lower \gls{bler} than conventional \gls{noma} and cooperative \gls{oma} \cite{8695100}. However, multi-antenna \gls{noma} is shown to be an inefficient strategy in terms of multiplexing gain and use of \gls{sic} receivers in downlink multi-user multi-antenna systems \cite{9451194}.
\par Recently, \gls{rsma} has emerged as a novel multiple access for downlink multi-antenna networks under infinite blocklength assumption. By splitting user messages, \gls{rsma} can softly bridge two extreme interference management strategies, namely, \gls{noma} and \gls{sdma} \cite{Mao2018,8907421}.
\gls{rsma} offers significant gains in terms of energy efficiency, \gls{csi} feedback overhead reduction and user fairness \cite{9831440}.
With \gls{fbl} constraints, it was demonstrated that RSMA could guarantee the same transmission rate while allowing a reduction in blocklength and therefore reducing the transmission delay compared to \gls{sdma} and \gls{noma} \cite{9831048}.
\par User fairness is an important criterion, however, to the best of our knowledge, user fairness of \gls{rsma} under \gls{fbl} assumption has not been investigated. The user rate is composed of private rate and common rate, while the common rate is constrained by the rate of the worst-case user since the common stream must be decoded by all users. The common rate may decrease when users experience heterogeneous downlink channel strength. One possible solution to the problem is combining \gls{rsma} with cooperative transmission utilizing user relaying. We also consider a complex scenario involving simultaneous transmitting distinct messages to multiple multicast groups, which is likely to occur in the future wireless network due to the content-based services \cite{8019852}.

Compared to the previous work \cite{8846761}, we propose a novel system model that unifies the multigroup multicast and the cooperative transmission models, and we extend the proposed model to \gls{fbl} scenarios. We study the \gls{mmf} rate of \gls{rsma} with/without cooperative transmission under the constraints on blocklength and transmit power in both underloaded and overloaded deployments.
Our results show that by utilizing \gls{rsma}, the user fairness can be guaranteed more efficiently compared to power-domain \gls{noma}\footnote{In the sequel, power-domain NOMA will be referred simply by \gls{noma}.} and \gls{sdma} with the same \gls{fbl}. Alternatively, the blocklength can be reduced, and the latency is therefore decreased while achieving the same \gls{mmf} rate. With cooperative transmission, the gain of \gls{rsma} can be enhanced further.
\par The rest of the paper is organized as follows. In Section \ref{sec:system_model}, we introduce the system model of multigroup multicast \gls{rsma} with/without user relay and we formulate \gls{mmf} optimization problem. The proposed \gls{1dsca} joint optimization algorithm is specified in Section \ref{sec:proposed_algorithm}. Numerical results illustrating the benefits of \gls{rsma} and Cooperative \gls{rsma} (C-RSMA) are discussed in Section \ref{sec:results_and_discussion}, followed by the conclusions in Section \ref{sec:conclusion}.
\par The superscript $(\cdot)^T$ denotes transpose and $(\cdot)^H$ denotes conjugate-transpose. $\mathcal{CN}(\zeta,\varphi^2)$ represents a complex Gaussian distribution with mean $\zeta$ and variance $\varphi^2$. The boldface uppercases represent matrices and lowercase letters represent vectors. $\text{tr}(\cdot)$ is the trace. $|\cdot|$ is the absolute value and $\|\cdot\|$ is the Euclidean norm. $\mathbb{C}$ denotes the complex space. $|\mathcal{A}|$ is the cardinality of the set $\mathcal{A}$.

\section{System model and problem formulation}\label{sec:system_model}
\par We design a system model that unifies the multigroup multicast and the cooperative transmission models. We assume that a \gls{bs} is equipped with $N_t$ transmit antennas communicating with $K$ single-antenna users that are indexed by $\mathcal{K}=\{1,2,\ldots,K\}$. The users are divided into $M\in[1,K]$ separated groups $\{\mathcal{G}_1,\ldots,\mathcal{G}_M\}$, where $\mathcal{G}_m\subseteq \mathcal{K}$ is the set of users belonging to the $m$th group, $m\in\mathcal{M}$, and $\mathcal{M}\triangleq\{1,\ldots,M\}$ is the index set of all groups.
The user grouping is done such that the groups satisfy $\bigcup_{m\in\mathcal{M}}\mathcal{G}_m=\mathcal{K}$ and $\mathcal{G}_m\cap\mathcal{G}_j=\emptyset,\,\forall m,j\in\mathcal{M},m\neq j$.
The mapping function $\mu(k)=m$ maps a user-$k$ to group-$m$.
The \gls{bs} sends the distinct messages $W_1,\ldots,W_M$ to users in $\mathcal{G}_1,\ldots,\mathcal{G}_M$, respectively, and users belonging to the same group, $\mathcal{G}_m$, require a same message $W_m$.
We denote the size of the group as $|\mathcal{G}_m|=G_m$.
\par
\begin{figure}[t!]
    \centering
    \hspace*{0.1cm}\includegraphics[scale=0.32]{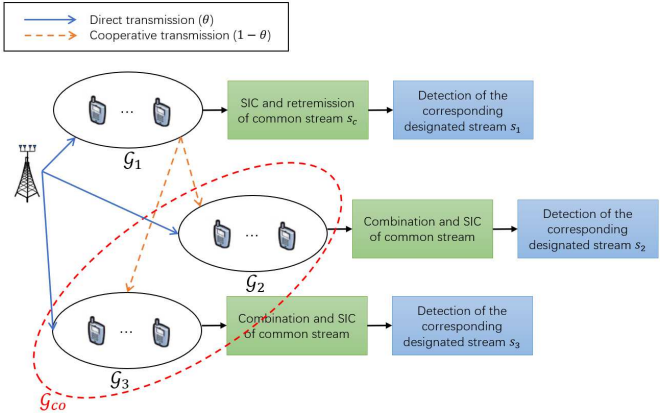}\vspace{-6mm}
    \caption{Unified system model with $M=3$.\vspace{-7mm}}
    \label{fig:sys_illu}
\end{figure}

 Fig. \ref{fig:sys_illu} shows the system model with $M=3$ user groups. The transmission consists of two phases, namely, the direct transmission (in the first time slot) and cooperative transmission phases (in the second time slot). Unlike \cite{8695100} that allocates equal transmission time to two phases, we consider a dynamic time allocation strategy. $\theta$ is the fraction of time allocated to the direct transmission phase. 
In the first time slot, signals are transmitted from \gls{bs} to all groups. In the second time slot, the user in $\mathcal{G}_1$ cooperatively decode and forward the signals to the users in $\mathcal{G}_2$ and $\mathcal{G}_3$.

\subsection{Direct transmission phase}
\par We consider an \gls{rsma}-assisted transmission model where the \gls{bs} splits the message $W_m$ for users in group-$m$ into $\{W_{c,m},W_{p,m}\}$, with $W_{c,m}$ and $W_{p,m}$ being the common and private parts, respectively. All common parts are combined into one common message and encoded into common stream {$s_{c}$} to be decoded by all users. The private message $W_{p,m}$ is encoded into $s_m$, and is only decoded by users in group-$m$. The overall symbol streams to be transmitted is denoted by $\mathbf{s}={[s_{c},s_1,s_2,\ldots,s_M]}^T$, and $\mathbb{E}[\mathbf{s}\mathbf{s}^H]=\mathbf{I}$. The streams are precoded via a precoding matrix $\mathbf{P}=[\mathbf{p}_{c},\mathbf{p}_1,\mathbf{p}_2, \ldots,\mathbf{p}_M]$, where $\mathbf{p}_j\in\mathbb{C}^{N_t\times1}$ represents the linear precoder for the stream $s_j$, $j\in\{c,1,2, \ldots,M\}$. This yields the transmit signal
\begin{equation}
    \label{Equ:Input_1_time_slot}
    \setlength{\abovedisplayskip}{3pt}
    \setlength{\belowdisplayskip}{3pt}
    \mathbf{x}^{[1]}=\mathbf{p}_cs_c+\sum\nolimits_{m\in\mathcal{M}}\mathbf{p}_ms_m,
\end{equation}
where the superscript $[1]$ represents the first transmission phase. The power constraint is $\mathrm{tr}(\mathbf{P}\mathbf{P}^H)\leq P_t$. Denote the signal received at the $k$th user in the $m$th group as $y_k$
\begin{equation}
    \label{Equ:output_signal}
    \setlength{\abovedisplayskip}{3pt}
    \setlength{\belowdisplayskip}{3pt}
    y_k=\mathbf{h}_k^H\mathbf{x}^{[1]}+n_k,
\end{equation}
where $\mathbf{h}_k\in\mathbb{C}^{N_t\times1}$ is the channel vector between the transmitter and the $k$th user, and $n_k\sim\mathcal{CN}(0,1)$ is the receiver \gls{awgn}. After each user receives the signal, the common stream is decoded and eliminated from the received signal, while the private parts are treated as noise. The \gls{sinr} of the common stream is
\begin{equation}
    \setlength{\abovedisplayskip}{3pt}
    \setlength{\belowdisplayskip}{3pt}
    \gamma_{c,k}^{[1]}=\frac{{|\mathbf{h}_k^H\mathbf{p}_{c}|}^2}{\sum_{j\in\mathcal{M}}{|\mathbf{h}_k^H\mathbf{p}_j|}^2+1}, \forall k\in\mathcal{K}.
\end{equation}
We assume that the transmitter and receivers perfectly know the channel vectors and perfect \gls{sic} can be performed at receivers. Once the common stream $s_c$ is successfully decoded, it is reconstructed and removed from the received signal. Subsequently, each user in $\mathcal{G}_m$ decodes the  group private stream while treating other private streams as noise.
The \gls{sinr} of decoding the private stream $s_m$ at user-$k$ is
\begin{equation}
    \setlength{\abovedisplayskip}{3pt}
    \setlength{\belowdisplayskip}{3pt}
    \gamma_{p,k}^{[1]}=\frac{{|\mathbf{h}_k^H\mathbf{p}_{m}|}^2}{\sum_{j\in\mathcal{M},j\neq m}{|\mathbf{h}_k^H\mathbf{p}_j|}^2+1}, \forall k\in\mathcal{G}_m.
\end{equation}
\par Once the common and private messages are decoded, each user in $\mathcal{G}_m$ reconstructs the original message by extracting the decoded common message and combining it with the decoded $W_{p,m}$. 
We include the effect of blocklength by using the rate expression proposed in \cite{5452208}. Denote the total blocklength during the transmission as $l_n=l_d+l_c$, where $l_d$ is blocklength allocated to the direct transmission phase and  $l_c$ is the blocklength of the codeword in the cooperative transmission phase. The rates $R_{c,k}^{[1]}, R_{p,k}^{[1]}$ of decoding the common and private streams (respectively denoted as common rate and private rate) in the first time slot are expressed as
\begin{equation}\label{Equ:total_Rate}
    \setlength{\abovedisplayskip}{3pt}
    \setlength{\belowdisplayskip}{3pt}
    R_{i,k}^{[1]} \approx \theta\left[log_2(1+\gamma_{i,k}^{[1]})-\sqrt{\frac{V(\gamma_{i,k}^{[1]})}{l_d}}B\right],\, i\in\{c,p\},
\end{equation}
where $B=Q^{-1}(\epsilon)\log_2e$, $\epsilon$ represents \gls{bler}, and $Q^{-1}(\cdot)$ corresponds to the inverse of the Gaussian $\mathcal{Q}$ function, $Q(x)=\int_x^{\infty}\frac{1}{\sqrt{2\pi}}\exp(-\frac{t^2}{2})dt$.  
$\theta=l_d/l_n$, since the blocklength of the codeword is proportional to the latency and can be approximately expressed as $l_n\approx BT$, where $B$ and $T$ are the bandwidth and time duration of the signal (i.e., latency), respectively \cite{7529226}. 
$V(\cdot)$ is the channel dispersion parameter
\begin{equation}\label{Equ:ChannelDisper}
    \setlength{\abovedisplayskip}{3pt}
    \setlength{\belowdisplayskip}{3pt}
    V(\gamma_{i,k}^{[1]}) = 1-\frac{1}{{(1+\gamma_{i,k}^{[1]})}^{2}}.
\end{equation}

\subsection{Cooperative transmission phase}
\par We assume that users in $\mathcal{G}_1$ are cell-center users. They employ the \gls{ndf} protocol and act as \gls{df} \gls{hd} relays. Users in $\mathcal{G}_1$ re-encode the decoded common stream $s_c$ using a different codebook generated independently from that of the \gls{bs}, and forward $s_c$ to users in other groups (denoted as $\mathcal{G}_{co}=\{\mathcal{G}_2,\ldots,\mathcal{G}_M\}$) with transmit power $P_r$. The transmit signal at user-$j$ of $\mathcal{G}_1$ in the second time slot (denoted as superscript $[2]$) is given by $x_j^{[2]}=\sqrt{P_r}s_c, \forall j\in\mathcal{G}_1$. The signal received at user-$k$ of $\mathcal{G}_{co}$ is 
\begin{equation}
    \setlength{\abovedisplayskip}{3pt}
    \setlength{\belowdisplayskip}{3pt}
    y_k^{[2]}=\sum\nolimits_{j\in\mathcal{G}_1}h_{k,j}x_j^{[2]}+n_k, \forall k\in\mathcal{G}_{co},
\end{equation}
where $h_{k,j}$ is the \gls{siso} channel between user-$k$ and user-$j$. The \gls{sinr} of decoding the common stream at user-$k$, $k\in\mathcal{G}_{co}$ is given by
\begin{equation}
    \setlength{\abovedisplayskip}{3pt}
    \setlength{\belowdisplayskip}{3pt}
    \gamma_{c,k}^{[2]}={\sum\nolimits_{j\in\mathcal{G}_1}{P_r|h_{k,j}|}^2}, \forall k\in\mathcal{G}_{co}.
\end{equation}
The corresponding rate is
\begin{equation}\label{equ:coop_rate_expres}
    \setlength{\abovedisplayskip}{3pt}
    \setlength{\belowdisplayskip}{3pt}
    R_{c,k}^{[2]} \approx (1-\theta)\left[\log_2(1+\gamma_{c,k}^{[2]})-\sqrt{\frac{V(\gamma_{c,k}^{[2]})}{l_c}}B\right], \forall k\in\mathcal{G}_{co}.
\end{equation}
\par Users in $\mathcal{G}_{co}$ combine the decoded common stream in both time slots, the achievable rate of the common stream is
\begin{equation}
    \setlength{\abovedisplayskip}{3pt}
    \setlength{\belowdisplayskip}{3pt}
    R_c=\min\{R_{c}^{[1]},R_{c}^{[2]}\},
\end{equation}
where $R_{c}^{[1]}=\min_{k\in\mathcal{G}_1}\{R_{c,k}^{[1]}\}$ and $R_{c}^{[2]}=\min_{k\in\mathcal{G}_{co}}\{R_{c,k}^{[1]}+R_{c,k}^{[2]}\}$ are the achievable common rate of users in $\mathcal{G}_1$ and $\mathcal{G}_{co}$, respectively. $R_{c}$ 
guarantees that each user is able to correctly decode the common stream \cite{9123680,8846761}. We define $R_{c}=\sum_{m\in\mathcal{M}}C_m$, where $C_m$ is the rate at which $W_{c,m}$ is communicated, and $\mathbf{c}=[C_1,C_2, \ldots,C_M]$. Once $s_c$ is decoded and removed from the received signal, user-$k$ decodes the intended private stream. Accordingly, the rate of group-$m$ is
\begin{equation}
    \setlength{\abovedisplayskip}{3pt}
    \setlength{\belowdisplayskip}{3pt}
    R_{m}=C_{m}+\min_{k\in\mathcal{G}_m}R_{p,k}^{[1]},\,m\in\mathcal{M}.
\end{equation}

\par Since \gls{rsma} and \gls{noma} employ interference cancellation which may cause error propagation, we set the \gls{bler} threshold of \gls{rsma} and \gls{noma} to $\epsilon^{\text{RSMA}}=\epsilon^{\text{NOMA}}=5\times 10^{-6}$ so as to guarantee the approximated overall \gls{bler} is not larger than $10^{-5}$. The \gls{bler} of \gls{sdma} is $\epsilon^{\text{SDMA}}=10^{-5}$.
\begin{remark}
    If $\theta=1$, the cooperative transmission phase is turned off and the system model becomes a conventional multigroup multicast case without user relaying. If $0<\theta<1$ and each user group has a single user, the system model reduces to a $K$-user cooperative transmission scenario.
\end{remark}

\vspace{-3mm}
\subsection{Problem formulation}\label{sec:problem_formulation}
\vspace{-1mm}
The user fairness issue is the main focus of this work. In order to maximize the minimum group rate, the precoder $\mathbf{P}$, the common rate allocation $\mathbf{c}$, and the blocklengths, $l_d$ and $l_c$, are jointly optimized. The \gls{mmf} problem is formulated as
\begin{subequations}\label{Prob:coop_orig_prob}
    \setlength{\abovedisplayskip}{3pt}
    \setlength{\belowdisplayskip}{3pt}
    \begin{align}
        \max_{\mathbf{P},\mathbf{c},l_d, l_c} \quad
         & \min_{m\in\mathcal{M}}\,R_{m}                                                    \\
        \mbox{s.t.}\quad
         & \sum_{m^{\prime}\in\mathcal{M}}C_{m^{\prime}}\leq R_{c}\label{k_user_example_c1} \\
         & \mathrm{tr}(\mathbf{P}\mathbf{P}^{H})\leq P_t\label{k_user_example_c2}           \\
         & \mathbf{c}\geq\mathbf{0}\label{k_user_example_c3}.
    \end{align}
\end{subequations}
The common stream decodability is guaranteed by constraint (\ref{k_user_example_c1}). (\ref{k_user_example_c2}) is the transmit power constraint.

\section{Proposed algorithm}\label{sec:proposed_algorithm}
\par We propose a \gls{1dsca} algorithm to solve problem (\ref{Prob:coop_orig_prob}). One can notice that $l_d=l_n-l_c$. Hence for a given $l_n$ and a fixed value of $l_c$, $l_d$ can be obtained. We first fix the value of $l_c$, and then we use \gls{sca} to solve the non-convex problem. Once the optimal precoder for the fixed $l_c$ is attained, we increase $l_c$ and solve the problem iteratively. The blocklength and precoder corresponding to the highest \gls{mmf} are selected as optimal blocklength and precoder for the given blocklength $l_n$. 
For a given blocklength $l_n$ and fixed $l_c$ (hence, $l_d$ and $\theta$ are fixed, due to the relation $\theta=l_d/l_n$), the problem (\ref{Prob:coop_orig_prob}) can be transformed to
\begin{subequations}\label{Prob:coop_orig_prob_fixed_bl}
    \setlength{\abovedisplayskip}{3pt}
    \setlength{\belowdisplayskip}{3pt}
    \begin{align}
        \max_{\mathbf{P},\mathbf{c}} \quad
         & \min_{m\in\mathcal{M}}\,R_{m}                                                      \\
        \mbox{s.t.}\quad
         & (\ref{k_user_example_c1}),\,(\ref{k_user_example_c2}),\,(\ref{k_user_example_c3}).
    \end{align}
    \vspace{-5mm}
\end{subequations}

\par Problem (\ref{Prob:coop_orig_prob_fixed_bl}) is non-convex due to the non-convex rate expressions. We introduce the following variables, namely $t$, $\boldsymbol{\alpha}_c=[\alpha_{c,1},\alpha_{c,2}, \ldots,\alpha_{c,K}]$, $\boldsymbol{\alpha}_p=[\alpha_{p,1},\alpha_{p,2}, \ldots,\alpha_{p,M}]$, $\boldsymbol{\rho}_c=[\rho_{c,1},\allowbreak\rho_{c,2},\allowbreak\ldots,\rho_{c,K}]$, $\boldsymbol{\rho}_p=[\rho_{p,1},\rho_{p,2}, \ldots,\rho_{p,K}]$. $t$ is the achievable rate lower bound for all user groups. $\theta\alpha_{p,m}$ is the lower bound of the minimum private rate of users in group-$m$, i.e., $\min_{k\in\mathcal{G}_m}{R_{p,k}^{[1]}}\geq\theta\alpha_{p,m}$. $\theta\alpha_{c,k}$ represents the lower bound of common rate $R_{c,k}^{[1]}$ in the direct transmission phase. $\rho_{c,k}$ and $\rho_{p,k}$ are the lower bounds of the \gls{sinr} of the common and private streams in the first phase, respectively. $\nu(\rho)=1-(1+\rho)^{-2}$.
Problem (\ref{Prob:coop_orig_prob_fixed_bl}) is equivalently written as

\begin{subequations}\label{Prob:2nd_transform}
    \setlength{\abovedisplayskip}{0pt}
    \setlength{\belowdisplayskip}{3pt}
    \begin{align}
        \max_{\substack{t,\mathbf{P},\mathbf{c},\boldsymbol{\alpha}_c,                                                                                                                                                             \\
                \boldsymbol{\alpha}_p,\boldsymbol{\rho}_c,\boldsymbol{\rho}_p}}
         & \quad\quad\quad t\label{2nd_transform_c0}                                                                                                                                                                               \\
        \mbox{s.t.}\quad\quad
         & (C_m+\theta\alpha_{p,m})\geq t,\,\forall m\in\mathcal{M}\label{2nd_transform_c1}                                                                                                                                        \\
         & \theta\alpha_{c,k}\geq\sum_{j\in\mathcal{M}}C_j,\,\forall k\in\mathcal{G}_1\label{2nd_transform_c2}                                                                                                                     \\
         & \theta\alpha_{c,k}+R^{[2]}_{c,k}\geq\sum_{j\in\mathcal{M}}C_j,\,\forall k\in\mathcal{G}_{co}\label{2nd_transform_c3}                                                                                                    \\
        \begin{split}
            &\log_2(1+\rho_{c,k}) - B\sqrt{\frac{\nu(\rho_{c,k})}{l_d}} \geq\alpha_{c,k},\,\forall k\in\mathcal{K}
        \end{split}\label{2nd_transform_c4}                                                                                                                          \\
        \begin{split}
            &\log_2(1+\rho_{p,k}) - B\sqrt{\frac{\nu(\rho_{p,k})}{l_d}} \geq \alpha_{p,\mu(k)},\,\forall k\in\mathcal{K}
        \end{split}\label{2nd_transform_c5}                                                                                                                    \\
         & \sum_{j\in\mathcal{M}}{|\mathbf{h}_k^H\mathbf{p}_j|}^2+1-\frac{{|\mathbf{h}_k^H\mathbf{p}_{c}|}^2}{\rho_{c,k}}\leq0,\,\forall k\in\mathcal{K}\label{2nd_transform_c6}                                                   \\
         & \hspace{-5mm} \sum_{\substack{j\in\mathcal{M},j\neq\mu(k)}}{|\mathbf{h}_k^H\mathbf{p}_j|}^2\!+\!1\!-\!\frac{{|\mathbf{h}_k^H\mathbf{p}_{\mu(k)}|}^2}{\rho_{p,k}}\leq0,\,\forall k\in\mathcal{K}\label{2nd_transform_c7} \\
         & (\ref{k_user_example_c2}), (\ref{k_user_example_c3}).
        \vspace{-4mm}
    \end{align}
\end{subequations}
\par Problem (\ref{Prob:2nd_transform}) remains non-convex due to the non-convex constraints (\ref{2nd_transform_c4})\verb|-|(\ref{2nd_transform_c7}). Next, we approximate the non-convex parts $\sqrt{\nu(\rho_{c,k})}$, $\sqrt{\nu(\rho_{p,k})}$ in the constraints by the first-order Taylor series. Constraints (\ref{2nd_transform_c4})\verb|-|(\ref{2nd_transform_c5}) and (\ref{2nd_transform_c6})\verb|-|(\ref{2nd_transform_c7}) are approximated at $\{\rho_{c,k}^{[n]}$, $\rho_{p,k}^{[n]}\}$ and $\{\mathbf{p}_{c}^{[n]}$, $\mathbf{p}_{k}^{[n]}\}$ at iteration \textit{n} as (\ref{equ:taylor_rate}) and (\ref{equ:sigma_expan}), respectively.
\begin{subequations}\label{equ:taylor_rate}
    \setlength{\abovedisplayskip}{3pt}
    \setlength{\belowdisplayskip}{3pt}
    \begin{align}
        \begin{split}
            &\log_2(1+\rho_{c,k})-\frac{B}{\sqrt{l_d}}\bigg\{\left[1-(1+\rho^{[n]}_{c,k})^{{-2}}\right]^{-\frac{1}{2}}\Big[(1+\rho^{[n]}_{c,k})^{-3}\\
                &\,\quad\quad(\rho_{c,k}-\rho^{[n]}_{c,k})-(1+\rho^{[n]}_{c,k})^{-2}+1\Big]\bigg\}\geq \alpha_{c,k},
        \end{split} \\
        \begin{split}
            &\log_2(1+\rho_{p,k})-\frac{B}{\sqrt{l_d}}\bigg\{\left[1-(1+\rho^{[n]}_{p,k})^{{-2}}\right]^{-\frac{1}{2}}\Big[(1+\rho^{[n]}_{p,k})^{-3}\\
                &\,\quad\quad(\rho_{p,k}-\rho^{[n]}_{p,k})-(1+\rho^{[n]}_{p,k})^{-2}+1\Big]\bigg\}\geq \alpha_{p,\mu{(k)}}.
        \end{split}
    \end{align}
\end{subequations}
\begin{subequations}\label{equ:sigma_expan}
    \setlength{\abovedisplayskip}{3pt}
    \setlength{\belowdisplayskip}{3pt}
    \begin{align}
         & \hspace{-2mm}  \sum_{j\in\mathcal{M}}{|\mathbf{h}_k^H\mathbf{p}_j|}^2\!+\!1\!-\!\frac{2\mathfrak{R}\{(\mathbf{p}_c^{[n]})^H\mathbf{h}_k\mathbf{h}_k^H\mathbf{p}_c\}}{\rho_{c,k}^{[n]}}\!+\!\frac{{|\mathbf{h}_k^H\mathbf{p}_c^{[n]}|}^2\rho_{c,k}}{(\rho_{c,k}^{[n]})^2}\!\leq\!0, \\
        \begin{split}
            &\hspace{-2mm} \sum_{\substack{j\in\mathcal{M},j\neq\mu(k)}}{|\mathbf{h}_k^H\mathbf{p}_j|}^2+1-\frac{2\mathfrak{R}\{(\mathbf{p}_{\mu{(k)}}^{[n]})^H\mathbf{h}_k\mathbf{h}_k^H\mathbf{p}_{\mu{(k)}}\}}{\rho_{p,k}^{[n]}}+\\
            &\,\quad\quad\quad\quad\quad\quad\quad\quad\frac{{|\mathbf{h}_k^H\mathbf{p}_{\mu{(k)}}^{[n]}|}^2\rho_{p,k}}{(\rho_{p,k}^{[n]})^2}\leq0.
        \end{split}
    \end{align}
\end{subequations}

\begin{figure*}[t!]
    \centering
    \floatsetup{valign=t, heightadjust=all}
    \ffigbox{%
        \begin{subfloatrow}
            \ffigbox[0.66\linewidth]{\includegraphics[scale=0.7]{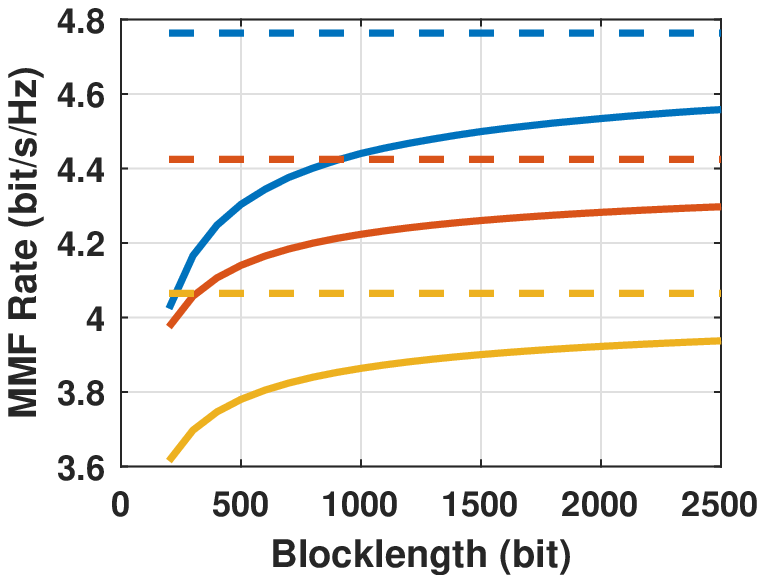}}{
                \caption{Underloaded deployment with $N_t=4$, $M=2$, $G_1=G_2=1$, $\varphi_1^2=1,\varphi_2^2=0.09$}
                \label{fig:2_Under_SNR20_1}}\hspace{-0.5mm}
            \ffigbox[0.66\linewidth]{\includegraphics[scale=0.7]{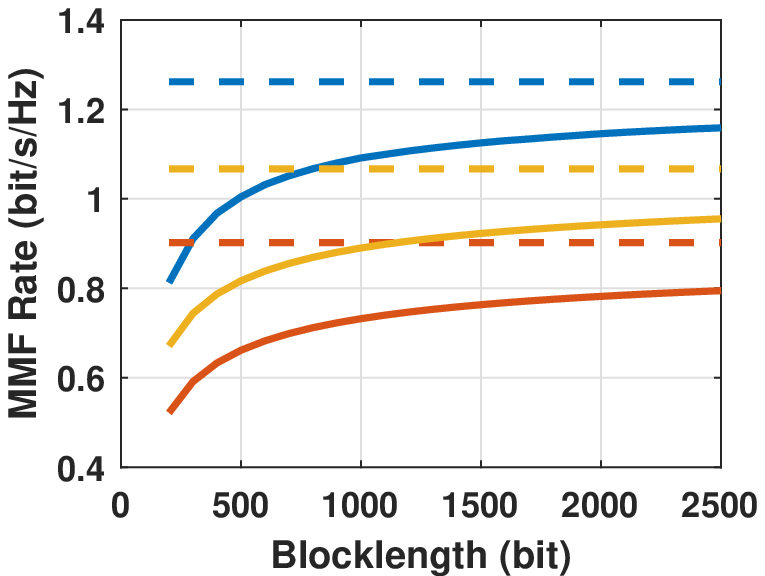}}{
                \caption{Overloaded deployment with $N_t=4$, $M=8$, $G_1=G_2=\ldots=G_8=1$,$\varphi_1^2=1$, $\varphi_2^2=0.875,\ldots,\varphi_8^2=0.125$}
                \label{fig:2_Over_2}}\hspace{2mm}
            \ffigbox[0.66\linewidth]{\includegraphics[scale=0.7]{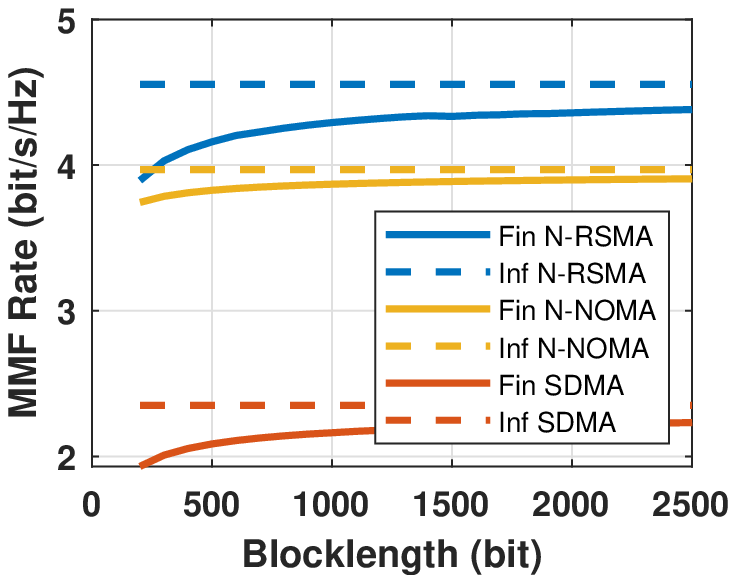}}{
                \caption{Overloaded deployment with $N_t=2$, $M=2$, $G_1=G_2=2$, $\varphi_1^2=\varphi_2^2=,\ldots,=\varphi_4^2=1$}
                \label{fig:2_multigroup_3}}
        \end{subfloatrow}}
    {\caption{\gls{mmf} rate versus blocklength of different strategies for non-cooperative multigroup multicast deployment with $\text{SNR}=20\text{dB}$.}
        \label{fig:2_user_3_SNR}}
\end{figure*}


Based on the approximation methods described above, the original non-convex problem is transformed to a convex problem and can be solved using the \gls{sca} approach. 
The main idea of \gls{sca} is to solve the non-convex problem by approximating it to a sequence of convex subproblems, which are solved successively. At iteration \textit{n}, based on the optimal solution ($\mathbf{P}^{[n-1]}$,$\boldsymbol{\rho}_c^{[n-1]}$, $\boldsymbol{\rho}_d^{[n-1]}$) obtained from the previous iteration $n-1$, we solve the following subproblem
\begin{subequations}\label{Prob:3rd_transform_final}
    \setlength{\abovedisplayskip}{3pt}
    \setlength{\belowdisplayskip}{3pt}
    \begin{align}
        \max_{\substack{t,\mathbf{P},\mathbf{c},\boldsymbol{\alpha}_c,            \\
        \boldsymbol{\alpha}_p,\boldsymbol{\rho}_c,\boldsymbol{\rho}_p}} \quad & t \\
        \mbox{s.t.}\quad
        \begin{split}
            &(\ref{equ:taylor_rate}), (\ref{equ:sigma_expan}),(\ref{k_user_example_c2}),(\ref{k_user_example_c3}).
        \end{split}
    \end{align}
\end{subequations}
The proposed \gls{1dsca} algorithm is summarized in Algorithm \ref{Algor:WSR}, where $\tau$ represents the tolerance of algorithm. 
Since the solution of Problem (\ref{Prob:3rd_transform_final}) at iteration $n-1$ is a feasible solution to the problem at iteration $n$, the convergence of Algorithm \ref{Algor:WSR} is ensured.
The objective variable $t$ is monotonically increasing and it is bounded above by the transmit power constraint.
We propose a different algorithm (namely, \gls{1dsca}) to solve the formulated problem instead of using \gls{wmmse} as in \cite{8846761} due to the non-convex FBL rate expressions in (\ref{Equ:total_Rate}) and (\ref{equ:coop_rate_expres}). The proposed algorithm is more general than the one in \cite{9831048} due to the optimization of the time allocation variable. By combining the one-dimensional search with \gls{sca}, less auxiliary variables and approximations are introduced and performance loss is hindered.
\par Algorithm \ref{Algor:WSR} consists of two loops, one is due to the one-dimensional search, and the other is due to the \gls{sca}-based algorithm. The worst-case computational complexity of the one-dimensional search algorithm is $\mathcal{O}(\delta^{-1})$, where $\delta\in(0,1)$ is the increment between two adjacent candidates of $\theta$. At each SCA iteration, the approximated problem is solved. Though additional variables $\boldsymbol{\alpha}_c,\boldsymbol{\alpha}_p,\boldsymbol{\rho}_c$ and $\boldsymbol{\rho}_p$ are introduced for convex relaxation, the main complexity still comes from the precoder design. The total number of SCA iterations required for the convergence is approximated as $\mathcal{O}\left(\log(\tau^{-1})\right)$. The worst-case computational complexity at each one-dimensional search iteration is $\mathcal{O}\left(\log(\tau^{-1})[KN_t]^{3.5}\right)$. Hence, the computational complexity of the proposed \gls{1dsca} is $\mathcal{O}\left(\delta^{-1}\log(\tau^{-1})[KN_t]^{3.5}\right)$.

\begin{algorithm}[t!]
    \caption{1D-SCA}\label{Algor:WSR}
    \textbf{Initialize}: $l_c=100$\;
    \Repeat{$l_d\leq100$}{
    \textbf{Initialize}: $n\leftarrow0$, $t^{[n]}\leftarrow0$, $\mathbf{P}^{[n]}$,$\boldsymbol{\rho}_c^{[n]}$, $\boldsymbol{\rho}_p^{[n]}$\;
    \Repeat{$|t^{[n]}-t^{[n-1]}|<\tau$}{
    $n\leftarrow n+1$\;
    {Solve problem (\ref{Prob:3rd_transform_final}) using $\mathbf{P}^{[n-1]}$,$\boldsymbol{\rho}_c^{[n-1]}$,$\boldsymbol{\rho}_p^{[n-1]}$ and denote the optimal value of the objective function as $t^{*}$ and the optimal solutions as $\mathbf{P}^{*}$,$\boldsymbol{\rho}_c^{*}$,$\boldsymbol{\rho}_p^{*}$}\;
    Update $t^{[n]}\leftarrow t^*$,$\mathbf{P}^{[n]}\leftarrow\mathbf{P}^{*}$,$\boldsymbol{\rho}_c^{[n]}\leftarrow\boldsymbol{\rho}_c^{*}$,$\boldsymbol{\rho}_p^{[n]}\leftarrow\boldsymbol{\rho}_p^{*}$\;
    }
    $l_c=l_c+10$\;
    }
\end{algorithm}

\section{Results and discussion}\label{sec:results_and_discussion}
This section investigates the MMF performance of \gls{rsma}, \gls{noma}, and \gls{sdma} for a variety of user deployments. Following the literature \cite{8846761,9123680,Mao2018}, the precoders of the proposed \gls{1dsca} algorithm are initialized by using \gls{mrt} combined with \gls{svd}. The tolerance of the algorithm is set to $\tau=10^{-3}$.
We first show the performance of a multigroup multicast model when $\theta$ is fixed to 1, then, we present the performance of a $K$-user cooperative model with optimized $\theta$. The problem (\ref{Prob:3rd_transform_final}) is solved using the CVX toolbox in Matlab \cite{grant2008cvx}.
\subsection{Multigroup multicast deployment}
Under the non-cooperative multigroup multicast deployment, $l_n=l_d$ and $\theta=1$.
Fig. \ref{fig:2_user_3_SNR} shows the \gls{mmf} performance of \gls{rsma}, \gls{noma} and \gls{sdma} with different number of transmit antennas and user deployments. $P_t=20$dB.
The \gls{mmf} performance is evaluated and averaged over $100$ random channel generations, where the channel $\mathbf{h}_k$ has \gls{iid} complex Gaussian entries with a certain variance, i.e. $\mathcal{CN}(0,\varphi_k^2)$.
In Fig. \ref{fig:2_Under_SNR20_1} and \ref{fig:2_Over_2}, we set the total number of users to $K=2$ and $K=8$ divided over $M=2$ and $M=8$ groups, respectively, such that each group contains one user. Fig. \ref{fig:2_multigroup_3} shows the performances under 2 transmit antennas and 4 users deployment. The users are divided into $M=2$ groups with 2 users per group.
Due to the complexity of finding the optimum decoding order in multigroup multicast deployment, we assume that the \gls{noma} \gls{sic} decoding order is performed in a descending order of channel gains in Fig. \ref{fig:2_Under_SNR20_1} and \ref{fig:2_Over_2}. In Fig. \ref{fig:2_multigroup_3}, we simplify \gls{noma} by restricting all precoder directions to be the same, as the problem (27) in \cite{8019852}, and optimize the shared precoder by the \gls{sca}-based algorithm.
\par The notations ``Inf" and ``Fin" in figures represent the schemes when $l_n=\infty$ and when $l_n$ is finite, respectively. Non-cooperative \gls{rsma} is marked as ``N-RSMA''.
In Fig. \ref{fig:2_user_3_SNR}, the \gls{mmf} rates with \gls{fbl} of three strategies increase with the blocklength as expected, and it is clear that with the same transmit power and blocklength, ``Fin N-RSMA” achieves a higher \gls{mmf} group rate. Compared with the underloaded deployment (in Fig. \ref{fig:2_Under_SNR20_1}), the relative gains of ``Fin N-RSMA" over ``Fin N-NOMA" and ``Fin SDMA" in overloaded scenario (in Fig. \ref{fig:2_Over_2}) are more pronounced as \gls{rsma} can manage the interference by partially decoding interference and partially treating interference as noise even in the overloaded deployment. ``Fin SDMA" achieves certain \gls{mmf} rate gain over ``Fin N-NOMA" in underloaded scenario while ``Fin N-NOMA" outperforms ``Fin SDMA" in the overloaded deployments, since \gls{sdma} can manage interference in the underloaded scenario and it cannot manage multi-user interference efficiently in the overloaded deployment.
From Fig. \ref{fig:2_multigroup_3}, the performance can be increased by 2.02 times by using \gls{rsma} compared with \gls{sdma} at the blocklength of 200, which is even higher than that in the infinite blocklength scenario.
By splitting group messages, \gls{rsma} achieves \gls{mmf} rate gain compared to \gls{noma} and \gls{sdma} regardless of user deployments and blocklength, which guarantees the user fairness in both infinite and finite scenarios. Alternatively, \gls{rsma} can utilize smaller blocklength (and hence lower latency) for achieving the same \gls{mmf} rate.

\subsection{$K$-user cooperative deployment}
\begin{figure}[t!]
    \centering
    \floatsetup{valign=t, heightadjust=all}
    \ffigbox{%
        \begin{subfloatrow}
            \ffigbox{\hspace*{-0.2cm}\includegraphics[scale=0.6]{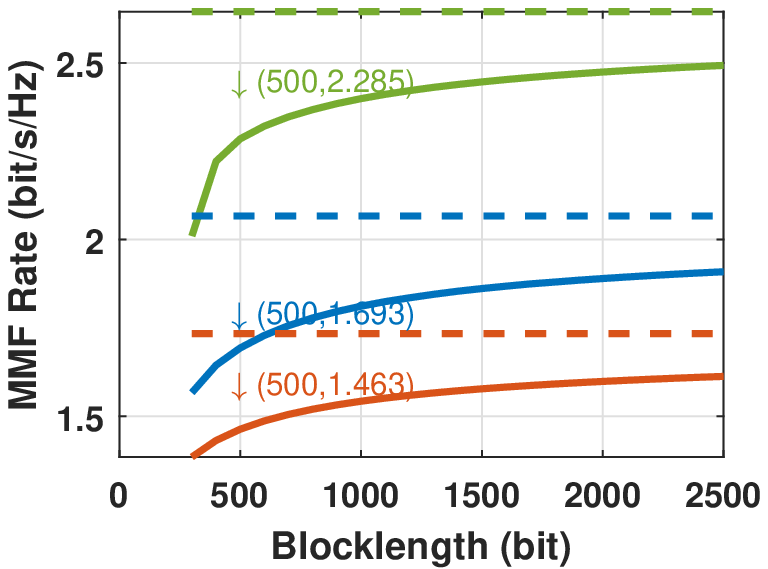}}{
                \caption{Underloaded scenario, $N_t=4$}
                \label{fig:coop_nt4}}
            \ffigbox{\hspace*{-0.15cm}\includegraphics[scale=0.6]{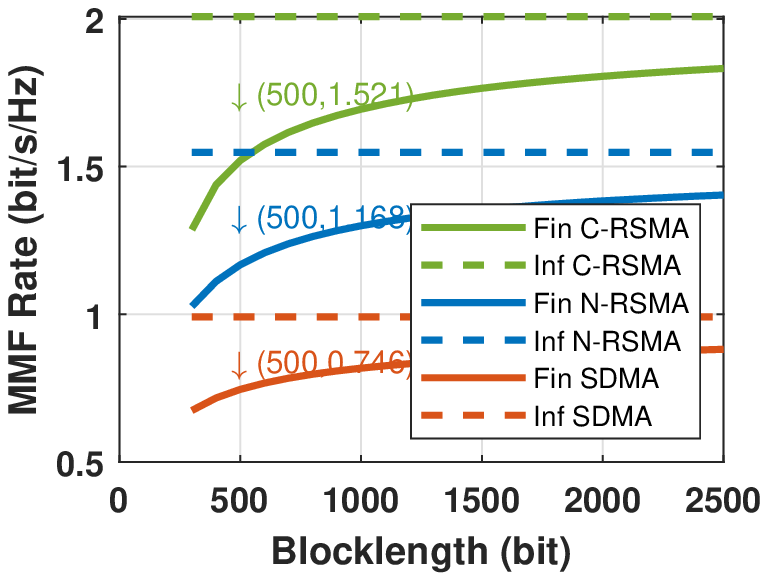}}{
                \caption{Overloaded scenario, $N_t=2$}
                \label{fig:coop_nt2}}
        \end{subfloatrow}}
    {\caption{\gls{mmf} rate performance versus blocklength for cooperative deployment, $M=3, G_1=G_2=G_3=1$, $\varphi_1^2=1$, $\varphi_2^2=0.09$, $\varphi_3^2=0.01$.}
        \label{fig:coop}}
\end{figure}

In this subsection, $\theta\in(0,1)$, the effects of cooperative transmission phase will be investigated. Fig. \ref{fig:coop} shows the average \gls{mmf} rates of different strategies 
with varied number of transmit antennas and channel strength disparities among users. The MMF values for blocklength 500 are given in the figure with arrows. We assume that $M=3,\,G_1=G_2=G_3=1$. The transmit power at relay-user is the same as that at \gls{bs}, $P_r=P_t$. The variances of user channels are $\varphi_1^2=1$,$\varphi_2^2=0.09$,$\varphi_3^2=0.01$. $N_t=4$ in subfigure \ref{fig:coop_nt4}, while $N_t=2$ in subfigure \ref{fig:coop_nt2}.

\par
\begin{figure}[t!]
    \centering
    \floatsetup{valign=t, heightadjust=all}
    \ffigbox{%
        \begin{subfloatrow}
            \ffigbox{\hspace*{-0.12cm}\includegraphics[scale=0.6]{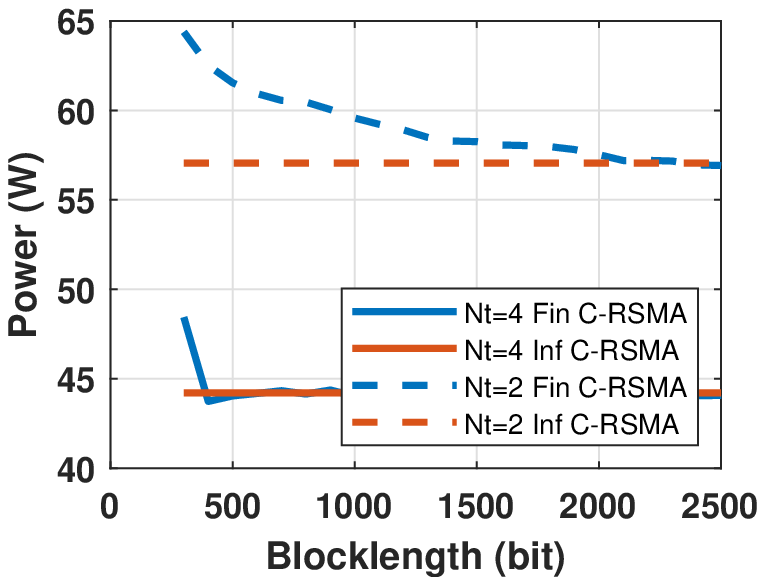}}{
                \caption{Power allocated to the common stream}
                \label{fig:coop_nt4_nt2_common_power}}
            \ffigbox{\hspace*{-0.1cm}\includegraphics[scale=0.6]{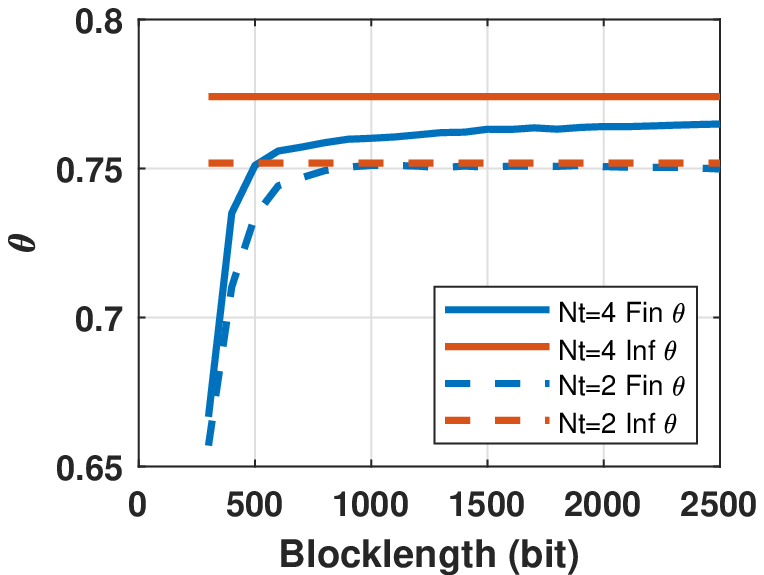}}{
                \caption{Time allocated to the direct transmission phase}
                \label{fig:coop_nt4_theta}}
        \end{subfloatrow}}
    {\caption{Time and power allocation in cooperative deployment.}
        \label{fig:applying_inf_precoder}}
\end{figure}

 From Fig. \ref{fig:coop_nt4} and \ref{fig:coop_nt2}, the performance of ``Fin C-RSMA'' outperforms ``Fin N-RSMA'' and ``Fin SDMA''. Furthermore, in the underloaded scenario, the ``Fin N-RSMA'' is likely to turn off the common message and boil down to \gls{sdma}, especially when blocklength is small, resulting the relative gain (N-RSMA over \gls{sdma}) close to $0$. But the trend is reversed through implementing cooperative transmission, and the \gls{mmf} rate of \gls{rsma} attains nearly twice as much as that of \gls{sdma}. The system is overloaded when $N_t=2$. Compared to the underloaded deployment, when blocklength is 500 bits the relative gains of ``Fin C-RSMA'' over ``Fin SDMA'' and ``Fin N-RSMA'' over ``Fin SDMA'' are enhanced from 0.55 to 1 and from 0.14 to 0.55, respectively.
This is because each user experiences more severe multi-user interference in the overloaded scenario than in the underloaded case. In comparison to underloaded scenarios, a large portion of users' messages are split and encoded into the common stream for each user to decode, and more power is allocated to the common stream,
as shown in Fig. \ref{fig:coop_nt4_nt2_common_power}.
The amount of interference that will be decoded at each user is further increased as the relaying user in C-RSMA retransmits the common stream.

\par Fig. \ref{fig:coop_nt4_theta} shows how the time allocated to direct transmission phase, $\theta$, changes with blocklength for C-RSMA strategy. As depicted in Fig. \ref{fig:coop_nt4_theta}, less time is allocated to the direct transmission phase when the blocklength is small, and as blocklength increases, the gap between ``Inf $\theta$'' and ``Fin $\theta$'' becomes smaller. Since \gls{fbl} results in rate penalty, in order to improve user fairness, more time is allocated to the cooperative transmission phase to increase the \gls{mmf} rate of the cell edge users. C-RSMA with \gls{fbl} is better suited to situations where users experience stronger multi-user interference as it has a higher capability to manage interference.
Compared with N-RSMA and SDMA, C-RSMA achieves a higher \gls{mmf} rate with the same blocklength or attains the same \gls{mmf} rate with smaller blocklength (and therefore lower latency).
\par
\begin{figure}[t!]
    \centering
    \includegraphics[scale=0.7]{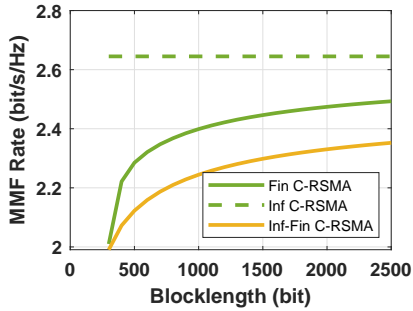}
    \caption{Applying the precoder from infinite blocklength scenario to FBL scheme in underloaded cooperative deployment.\vspace{-5mm}}
    \label{fig:coop_nt4_apply_inf_CRS}
\end{figure}

 We analyze the \gls{mmf} rate achieved by applying the precoder obtained under the infinite blocklength assumption to the scenario with \gls{fbl}, i.e., “Inf-Fin C-RSMA”.
From Fig. \ref{fig:coop_nt4_apply_inf_CRS}, a clear gap between “Fin RSMA” and “Inf-Fin RSMA” is observed, which justifies the effectiveness of our optimization.

\section{Conclusion}\label{sec:conclusion}
This paper investigates the performance of \gls{rsma} with \gls{fbl} in terms of user fairness in both non-cooperative and cooperative deployments. A novel system model that unifies non-cooperative/cooperative multigroup multicast deployments is proposed and an optimization problem is formulated to maximize the minimum group rate. A \gls{1dsca} algorithm is adopted to solve the problem. The results show that \gls{rsma} can better maintain user fairness with the same \gls{fbl} or achieve the same \gls{mmf} rate with smaller blocklength (and hence lower latency) in comparison with \gls{noma} and \gls{sdma}. With cooperative transmission, the gain of \gls{rsma} is enhanced further. Consequently, we conclude that \gls{rsma} is a promising strategy for enhancing user fairness in \gls{fbl} communications.

\vspace{-2mm}
\bibliographystyle{IEEEtran}
\bibliography{IEEEabrv,ref_test_CollegePC}

\end{document}